\documentclass{ws-p8-50x6-00}
\usepackage{overcite}
\usepackage{amssymb}
\usepackage{pennames}
\def\beq{\begin{equation}}
\def\eeq{\end{equation}}
\def\beqs{\begin{equation*}}
\def\eeqs{\end{equation*}}
\def\beqa{\begin{eqnarray}}
\def\eeqa{\end{eqnarray}}
\def\beqas{\begin{eqnarray*}}
\def\eeqas{\end{eqnarray*}}
\def\bit{\begin{itemize}}
\def\eit{\end{itemize}}
\def\ben{\begin{enumerate}}
\def\een{\end{enumerate}}
\newcommand{\bec}{\begin{center}}
\newcommand{\eec}{\end{center}}
\newcommand{\bce}{\begin{center}}
\newcommand{\ece}{\end{center}}
\def\ifmath#1{\relax\ifmmode #1\else $#1$\fi}%
\def\eg{{\it e.g.}}%
\def\etal{{\it et~al.}}%
\def\ie{{\it i.e.}}%
\def\PZ{\ifmath{\mathrm{Z}}}%
\def\simgt{\ifmath{\,\hbox{\lower0.6ex\hbox{$\sim$}\llap{\raise0.6ex\hbox{$>$}}}\,}}
\def\simlt{\ifmath{\,\hbox{\lower0.6ex\hbox{$\sim$}\llap{\raise0.6ex\hbox{$<$}}}\,}}
\let\muu=\mu
\def\mu{\ifmath{\muu}}%
\let\tauu=\tau
\def\tau{\ifmath{\tauu}}%
\let\psii=\psi  %  Save normal "\psi" definition, since I redefine it.
\def\psi{\ifmath{\psii}}%
\let\pii=\pi
\def\pi{\ifmath{\pii}}%
\def\eV{\hbox{\ifmath{\mathrm{e\kern-0.1em V}}}}%
\def\eVc{\hbox{\ifmath{\mathrm{e\kern-0.1em V\kern-0.16em}/\kern-0.1em\ifmath{c}}}}%
\def\eVcsq{\hbox{\ifmath{\mathrm{e\kern-0.1em V\kern-0.16em}/\kern-0.1em\ifmath{c^2}}}}%
\def\TeV{\hbox{\ifmath{\mathrm{T}}\kern-0.08em\eV}}%
\def\TeVc{\hbox{\ifmath{\mathrm{T}}\kern-0.1em\eVc}}%
\def\TeVcsq{\hbox{\ifmath{\mathrm{T}}\kern-0.1em\eVcsq}}%
\def\GeV{\hbox{\ifmath{\mathrm{G}}\eV}}%
\def\itGeV{\hbox{\it Ge\kern-1.2pt V}}%
\def\itMeV{\hbox{\it Me\kern-1.2pt V}}%
\def\itkeV{\hbox{\it ke\kern-1.2pt V}}%

\newbox\boxsqbox
\newdimen\boxsize\boxsize=1.2ex%
\def\boxop{%
\setbox\boxsqbox=\vbox{\hrule depth0.8pt width0.8\boxsize height0pt%
                       \kern0.8\boxsize
                       \hrule height0.8pt width0.8\boxsize depth0pt}%
           \hbox{%
           \vrule height1.0\boxsize width0.8pt depth0pt%
           \copy\boxsqbox
           \vrule height1.0\boxsize width0.8pt depth0pt\kern1.5pt}}%
\newcommand{\journal}[4]{\Journal{{#1}}{{#2}}{{#4}}{{#3}}}
\def\PLB{{\em Phys.\ Lett.}\ B}

\def\PRC{{\em Phys.\ Rev.}\ C}

\def\CPC{{\em Comp.\ Phys.\ Comm.}}
\newcommand {\EPJC}{{\em Eur.\ Phys. J.}\ C}

\newcommand{\JETSET}{{\sc jetset}}%

\renewcommand{\Re}{\ensuremath{\mathrm{Re}}}
\begin{document}

\title{New Bose-Einstein Results from L3}
\author{W. J. Metzger}

\address{University of Nijmegen, Toernooiveld 1, 6525 ED\ \ Nijmegen,
The Netherlands\\E-mail: W.Metzger@cern.ch}

%\address{{\rm (for the \Lthree\ collaboration)}}

%%%%%%%%%%%%%%%%%%%%%%%%%%%%%%%%%%%%%%%%%%%%%%%%%%%%%%%%%%%%%%
% You may repeat \author \address as often as necessary      %
%%%%%%%%%%%%%%%%%%%%%%%%%%%%%%%%%%%%%%%%%%%%%%%%%%%%%%%%%%%%%%

\maketitle

\abstracts{
In hadronic \PZ\ decays, we compare Bose-Einstein correlations in \Pgpz\ pairs
with those in identical charged pion pairs.  We also measure Bose-Einstein correlations
in triplets of identical charged pions.
From the first study, the source radius is found to be smaller for \Pgpz-\Pgpz\
than for \Pgppm-\Pgppm, as would be expected, \eg, in a string model.
However, the second study shows that the data are consistent with the pion production
being completely incoherent.
Both studies use data collected by the L3 collaboration at LEP.
}

%
%%%%%%%%%%%%%%%%%%%%%%%%%%%%%%%%%%%%%%%%%%%%%%%%%%%%%%%%%%%%%%%%%%%%%%%%%%%%%%%
% Introduction
%%%%%%%%%%%%%%%%%%%%%%%%%%%%%%%%%%%%%%%%%%%%%%%%%%%%%%%%%%%%%%%%%%%%%%%%%%%%%%%
%
\section{Introduction}
%While Bose-Einstein correlations (BEC) have been extensively studied for identical
%charged pions, this has rarely been the case for neutral pions.
%Three-particle BEC, although  occasionally studied, allow a
%test of whether the pions are incoherently produced, as is
%assumed in the simplest analyses of BEC.
%Both of these topics are investigated here using hadronic \PZ\ decays detected by L3 at LEP.

Two studies of Bose-Einstein correlations (BEC) are reported 
using hadronic \PZ\ decays detected by L3 at LEP.
The first uses two-particle BEC to compare the size of the emission region of \Pgpz\ to that of \Pgppm.
The second uses three-particle BEC to investigate the incoherence of pion emission.
Neither of these topics has previously been investigated in \Pep\Pem\ interactions.

\section{Comparison of BEC in \Pgpz-\Pgpz\ and \Pgppm-\Pgppm}

Since this work is completed and the paper accepted for publication,\cite{piz}
I only summarize it briefly here.
The analysis uses a data sample of about 2 million
hadronic \PZ\ decays corresponding to about 78 pb$^{-1}$.
Making use of the correlation function,
\begin{equation}
    R_2(Q) = \frac{\rho_2(Q)}{\rho_0(Q)}      \label{R2r}
\end{equation}
where $\rho_2$ is the number density of pion pairs for the data and
$\rho_0$ is the number density expected in the absence of BEC, both as a function
of the four-momentum difference, $Q$, between the two pions making up the pair.
Under the simplified assumptions of a static, spherical source with a Gaussian density, $R_2$
can be parametrized as
\begin{equation}   \label{R2param}
    R_2(Q) = {\cal N}(1+\alpha Q)(1+\lambda e^{-Q^2R^2})
\end{equation}
where ${\cal N}(1+\alpha Q)$ serves as normalization, taking into account some long-range correlation in $Q$.  
The factor $\lambda$ is equal to unity if pion production is incoherent and all pions come from the
hypothesized source.  The latter is certainly not the case because of long-lived resonances, and usually
$\lambda$ is found to be less than unity. 
With event and particle selection kept as identical as possible for \Pgpz-\Pgpz\
and \Pgppm-\Pgppm, and using \JETSET\cite{jetset} without any BEC simulation to determine $\rho_0$,
fits of (\ref{R2param}) result in a smaller radius for \Pgpz-\Pgpz\ than for \Pgppm-\Pgppm:
\begin{eqnarray*}
  R_{\pm\pm} - R_{00}   &=& 0.15\pm0.08\pm0.07 \ \mbox{fm}      \\
  R_{00}\ /\ R_{\pm\pm} &=& 0.67\pm0.16\pm0.15
\end{eqnarray*}

\section{Three-particle BEC}
This analysis is preliminary and at present makes use of a sample of about 1 million hadronic \PZ\ decays.
Similarly to the two-particle case, we use the correlation function,
\begin{equation}
    R_3(Q) = \frac{\rho(Q_3)}{\rho_0(Q_3)}
\end{equation}
where     $ Q_3^2={M_{123}^2-9m_\pi^2}=Q_{12}^2+Q_{23}^2+Q_{13}^2$ with
$Q_{ij}$ the four-momentum difference between the two pions $i$ and $j$ of the triplet.
There are contributions to $R_3$ from two-particle BEC and from genuine three-particle BEC.
With $G$ the Fourier transform of the source density,
%\begin{equation}
% R_3(Q_3) = 1 + \lambda
%             \left(|{ G(Q_{12})}|^2
%                 + |{ G(Q_{23})}|^2
%                 + |{ G(Q_{13})}|^2\right)
%              +    \lambda^{1.5}  \Re\{{ G(Q_{12})}
%                                       { G(Q_{23})}
%                                       { G(Q_{13})}\}
%\end{equation}
%The last term is the contribution to genuine three-particle BEC.
\begin{eqnarray}
 R_2(Q_{ij})          &=& 1 + \lambda   |{ G(Q_{ij})}|^2        \label{R2}         \\
 R_3^\mathrm{non-gen}(Q_3) &=& 1 + \lambda
             \left(|{ G(Q_{12})}|^2
                 + |{ G(Q_{23})}|^2
                 + |{ G(Q_{13})}|^2\right)                      \label{Rnongen}          \\
 R_3^\mathrm{gen}(Q_3) &=&
  1 + { 2{\lambda^{1.5}}\,\Re\{G(Q_{12})G(Q_{23})G(Q_{13})\}}   \label{Rgen}           \\
  R_3(Q_3) & = & R_3^\mathrm{non-gen}(Q_3) + R_3^\mathrm{gen}(Q_3) - 1  \label{R3}
\end{eqnarray}
Note that $R_3^\mathrm{gen}$ depends on the phase of $G$, whereas
$R_2$ and $R_3^\mathrm{non-gen}$ do not. With $G(Q_{ij})=|G|\exp(\imath\phi_{ij})$
and $\phi=\phi_{12}+\phi_{23}+\phi_{13}$, we see from (\ref{R2}) and (\ref{Rgen}) that
\begin{equation}
   \cos\phi(Q_3) = \frac{R_3^\mathrm{gen}(Q_3) - 1}
                        {2\sqrt{(R_2(Q_{12})-1)(R_2(Q_{23})-1)(R_2(Q_{13})-1)}}  \label{cosphi}
\end{equation}
Departure of $\cos\phi$ from unity occurs if the pion source is asymmetric of if the pion emission is
(partially) coherent. Genuine three-particle BEC have previously been observed in hadronic \PZ\
decay\cite{prev3}.  Our interest here is therefore mainly to investigate whether $\cos\phi=1$.
To measure $\cos\phi$, we must first measure $R_2$ using (\ref{R2r}), $R_3=\frac{\rho_3}{\rho_0}$ and
$R_3^\mathrm{non-gen}=\frac{\rho_1\rho_2}{\rho_0}$ from which $R_3^\mathrm{gen}$ is found using
(\ref{R3}). Here $\rho_n$ refers to the $n$-particle density and $\rho_0$ the corresponding density
of the `reference sample', \ie, a sample without BEC.
Then $\cos\phi$ is found from (\ref{cosphi}).

The reference sample is constructed using a mixing technique and corrected by Monte Carlo for non-BEC
correlations\cite{elong}.
Corrections for detector acceptance and efficiency, also determined from Monte Carlo, are applied to the
data. There are six different combinations of Monte Carlo programs used to determine these corrections.
Differences in results using the different combinations are the dominant contribution to the systematic
uncertainty.
Further, the $Q$-distributions are corrected for Coulomb repulsion by weighting each pair of pions by the
inverse Gamow factor\cite{gamow}.  At very small $Q$, statistics are small and the corrections are large;
accordingly the regions $Q_3<0.16 \,\GeV$ and $Q<0.08 \,\GeV$ are not used.

\begin{figure}[h]
\begin{center}
\epsfig{file=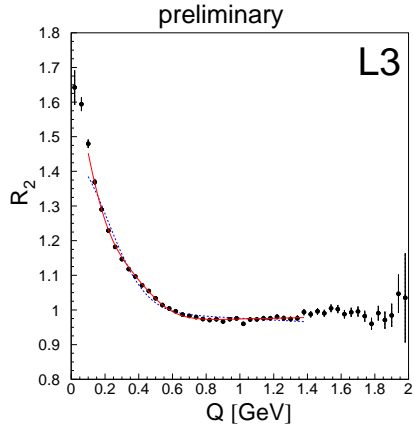,width=.46\textwidth,clip=}
\caption{Distribution of $R_2$.
The dashed (full) line is the result of a fit using the Gaussian (Edgeworth) parametrization.
}
\label{fig:R2}
\end{center}
\vspace{-5mm}
\end{figure}

\begin{figure}[ht]
\begin{center}
\epsfig{file=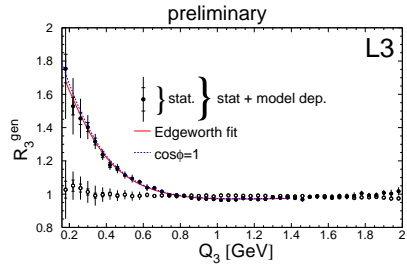,width=.5\textwidth,clip=}
\caption{Distribution of $R_3^\mathrm{gen}$. 
The full line is the result of a fit using the Edgeworth parametrization, while the dashed line is
the expectation from the Edgeworth fit to $R_2$ and the assumption $\cos\phi=1$. 
The open circles are the result of analyzing a MC sample without BEC as though it were data.
}
\label{fig:R3}
\end{center}
\vspace{-6mm}
\end{figure}

\begin{figure}[ht]
\begin{center}
\epsfig{file=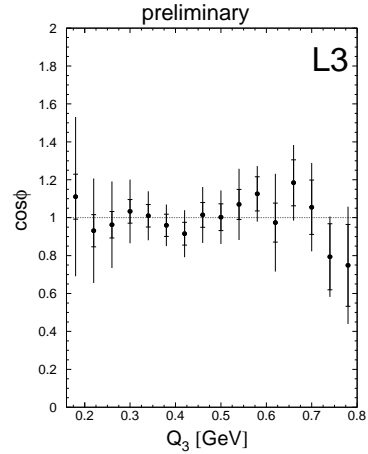,width=.4\textwidth,clip=,
        bbllx=30,bblly=49,bburx=653,bbury=840
}
\caption{Distribution of $\cos\phi$.
}
\label{fig:cos}
\end{center}
\vspace{-5mm}
\end{figure}

The resulting distributions of $R_2$ and $R_3^\mathrm{gen}$ are shown in Figs.\,\ref{fig:R2} and
\ref{fig:R3}, respectively.
As found in previous studies\cite{elong}, the Gaussian parametrization  (\ref{R2param}) of $R_2$
is found to be
inferior to a parametrization using the Edgeworth expansion about a Gaussian,
\begin{equation}   \label{R2Eparam}
    R_2(Q) = {\cal N}(1+\alpha Q)\left[1+\lambda e^{-Q^2R^2}
                                            ( 1+\kappa H_3(\sqrt{2}RQ) /6 )\right] 
\end{equation}
where $H_3$ is a 3$^\mathrm{rd}$ order Hermite polynomial and $\kappa$ a free parameter.
Also $R_3^\mathrm{gen}$ is found to be better fit by an Edgeworth expansion than by a Gaussian
parametrization. The result of this Edgeworth fit is shown in Fig.\,\ref{fig:R3}.
Also shown is the expectation calculated from the result of the fit to $R_2$
together with the assumption $\cos\phi=1$. 
It is seen that $R_3^\mathrm{gen}$ is well described by this assumption.  
The values of $\lambda$ and $R$ found in the Edgeworth fits are
 shown in Table~\ref{tab}.
%The values found from the fits to $R_2$ and to $R_3^\mathrm{gen}$ agree.
The values found in these fits agree.
Fig.\,\ref{fig:cos} shows the distribution of $\cos\phi$ found bin-by-bin from the Edgeworth
fit to $R_2$ and the distribution of $R_3^\mathrm{gen}$.  
It is seen that $\cos\phi$ is consistent with unity over the entire range of $Q_3$.

%\vfill
\begin{table}[htp]
\caption{Results of the Edgeworth fits to $R_2$ and $R_3^\mathrm{gen}$.
\label{tab}
}
\begin{center}
\begin{tabular}{|l|c|c|}\hline
 fit to                 & $\lambda$            & $R$ (fm)             \\ \hline
 $R_2$                  & $0.72\pm0.02\pm0.07$ & $0.73\pm0.02\pm0.04$ \\
 $R_3^\mathrm{gen}$ & $0.72\pm0.04\pm0.07$ & $0.72\pm0.04\pm0.05$ \\
\hline
\end{tabular}\end{center}
\end{table}
%\vfill

\noindent

%%%%%%%%%%%%%%%%%%%%%%%%%%%%%%%%%%%%%%%%%%%%%%%%%%%%%%%%%%%%%%%%%%%%%%%%%%%%%%%
% Conclusions 
%%%%%%%%%%%%%%%%%%%%%%%%%%%%%%%%%%%%%%%%%%%%%%%%%%%%%%%%%%%%%%%%%%%%%%%%%%%%%%%
%
\section{Conclusions}
The radius of the pion source found in BEC studies of
neutral pions is smaller than that of charged pions.
Studies of three-particle BEC show agreement
with completely incoherent
pion production.  Naively, the first result is expected in string models,
but the second result is surprising.

\twocolumn\noindent

%\end{multicols}
\section*{Acknowledgments}

The comparison of
BEC in \Pgpz-\Pgpz\ and \Pgppm-\Pgppm\ was done by M.P.~Sanders\cite{piz}
and the study of three-particle BEC by J.A.\,van Dalen.\cite{dalen}


\begin{thebibliography}{9}

\bibitem{piz}
%L3 Collab., P.~Achard \etal, \journal{\PLB}{}{ }{};
L3 Collab., P.~Achard \etal, {\PLB}, {accepted for publication};
M.P.~Sanders, Ph.D.\ thesis, Univ.\ of Nijmegen, 2002.

%\bibitem{L3det}
%L3 Collab., B.~Adeva \etal, \journal{\NIMA}{289}{1990}{35}; \\
%  J.A.~Bakken \etal, \journal{\NIMA}{275}{1989}{81};        \\
%  O.~Adriani \etal, \journal{\NIMA}{302}{1991}{53};         \\
%  B.~Adeva \etal, \journal{\NIMA}{323}{1992}{109};          \\
%  K.~Deiters \etal, \journal{\NIMA}{323}{1992}{162};        \\
%  B.~Acciari \etal, \journal{\NIMA}{351}{1994}{300};        \\
%  A.~Adam \etal, \journal{\NIMA}{383}{1996}{342}.

\bibitem{jetset}
%T. Sj{\"o}strand, \journal{\CPC}{39}{1986}{347};           \\
%T. Sj{\"o}strand and M. Bengtsson, \journal{\CPC}{43}{1987}{367}.
T. Sj{\"o}strand, \journal{\CPC}{82}{1994}{74}.

\bibitem{prev3}
DELPHI Collab., P.~Abreu \etal, \journal{\PLB}{355}{1995}{415};   %  \\
OPAL Collab., K.~Ackerstaff \etal, \journal{\EPJC}{5}{1998}{239}.

\bibitem{elong}
L3 Collab., M.~Acciarri \etal, \journal{\PLB}{458}{1999}{517}.

\bibitem{gamow}
M.~Gyulassy, S.~Kauffmann, and L.W.~Wilson, \journal{\PRC}{20}{1979}{2267}.


\bibitem{dalen}
%J.A.~van Dalen, W.~Kittel, and W.J.~Metzger, {\it Three-Particle Bose-Einstein Correlations in Hadronic Z
%Decay,} L3 Note ???? (2001);
J.A.~van Dalen, Ph.D.\ thesis, Univ.\ of Nijmegen, 2002.

%\bibitem{herwig}
%G.~Marchesini and B.~Webber, \journal{\NPB}{310}{1988}{461}; \\
%G. Marchesini \etal, \journal{\CPC}{67}{1992}{465}.

%\bibitem{tune}
%S.~Banerjee, D.~Duchesneau, S.~Sarkar, L3 Note 1818 (1995);  \\
%J.~Casaus, L3 Note 1946 (1996);                              \\
%L3 Collab., B.~Adeva \etal, \journal{\ZPC}{55}{1992}{39}.

%\bibitem{ariadne}
%L. L\"onnblad, \journal{\CPC}{71}{1992}{15}.
%%U.~Pettersson, {\em ARIADNE: A Monte Carlo for QCD Cascades in the Color Dipole
%%  Formulation}, Lund Preprint, LU TP 88-5 (1988);             \\
%%L.~L{\"o}nnblad, {\em The Colour Dipole Cascade Model and Ariadne Program}, Lund Preprint, LU TP 91-11
%%  (1991).


%\bibitem{Dreminpriv} I.M.~Dremin, private communication.


%\bibitem{dur}
%S.~Bethke, \etal, \journal{\NPB}{370}{1992}{310}.

%\bibitem{delphitalk}
%W.J.~Metzger, ``Oscillation of \Hq\ Moments as NOT a Test of QCD'',
%{\em Proc.~XXVIII Int.\ Symp.\ on Multiparticle Dynamics} (World Scientific, 1999);
%Nijmegen preprint HEN-417 (1998).

\end{thebibliography}
\end{document}

%%%%%%%%%%%%%%%%%%%%%%%%%%%%%%%%%%%%%%%%%%%%%%%%%%%%%%%%%%%%%%%%%%%%%%%%%%%%%
%% End of  ws-p8-50x6-00.tex  
%%%%%%%%%%%%%%%%%%%%%%%%%%%%%%%%%%%%%%%%%%%%%%%%%%%%%%%%%%%%%%%%%%%%%%%%%%%%%